# Superconductivity in Silicon Nanostructures

*N.T. Bagraev[1], W. Gehlhoff[2], L.E. Klyachkin[1], A.M. Malyarenko[1], and V.V. Romanov[1]*

[1]Ioffe Physico-Technical Institute, 194021 St. Petersburg, Russia
[2]Institut für Festkörperphysik, TU Berlin, D-10623 Berlin, Germany

**Abstract.** We present the findings of the superconductivity in the silicon nanostructures prepared by short time diffusion of boron after preliminary oxidation of the n-type Si (100) surface. These Si-based nanostructures represent the p-type high mobility silicon quantum well (Si-QW) confined by the $\delta$ - barriers heavily doped with boron. The ESR studies show that the $\delta$ - barriers appear to consist of the trigonal dipole centers, $B^+ - B^-$, which are caused by the negative-U reconstruction of the shallow boron acceptors, $2B^0 \rightarrow B^+ + B^-$. The temperature and magnetic field dependencies of the resistance, thermo-emf, specific heat and magnetic susceptibility demonstrate that the high temperature superconductivity observed seems to result from the transfer of the small hole bipolarons through these negative-U dipole centers of boron at the Si-QW – $\delta$-barrier interfaces. The value of the superconductor energy gap obtained is in a good agreement with the data derived from the oscillations of the conductance in normal state and of the zero-resistance supercurrent in superconductor state as a function of the bias voltage. These oscillations appear to be correlated by on- and off-resonance tuning the two-dimensional subbands of holes with the Fermi energy in the superconductor $\delta$-barriers. Finally, the proximity effect in the S-Si-QW-S structure is revealed by the findings of the multiple Andreev reflection (MAR) processes and the quantization of the supercurrent.

## 1. Introduction

Semiconductor silicon is well known to be the principal material for micro - and nanoelectronics. Specifically, the developments of the silicon planar technology are a basis of the metal-oxygen-silicon (MOS) structures and silicon-germanium (Si-Ge) heterojunctions that are successfully used as elements of modern processors (Macilwain 2005). Just the same goals of future high frequency processors especially to resolve the problem of quantum computing are proposed to need the application of the superconductor nanostructures that represent the Josephson junction series (Nakamura and Tsai 2000). Therefore the manufacture of superconductor device structures in frameworks of the silicon planar technology seems to give rise to new generations in nanoelectronics. Furthermore, one of the best candidate on the role of the superconductor silicon nanostructure appears to be the high mobility silicon quantum wells (Si-QW) of the p-type confined by the $\delta$-barriers heavily doped with boron on the n-type Si (100) surface which exhibit the properties of high temperature superconductors (Bagraev et al. 2006b). Besides, the heavily boron doping has been found to assist also the superconductivity in diamond (Ekimov et al. 2004). Here we present the findings of the electrical resistance, thermo-emf, specific heat and magnetic susceptibility measurements that are actually evidence of the superconductor properties for the δ-barriers heavily doped with boron which appear to result from the transfer of the small hole bipolarons through the negative-U dipole centres of boron at the Si-QW – δ-barrier interfaces. These 'sandwich' structures, S-Si-QW-S, are shown to be type II high temperature superconductors (HTS) with characteristics dependent on the sheet density of holes in the p-type Si-QW. The transfer of the small hole bipolarons appears to be revealed also in the studies of the proximity effect that is caused by the interplay of the multiple Andreev reflection (MAR) processes and the quantization of the supercurrent.

### 2. Sample preparation and analysis

The preparation of oxide overlayers on silicon monocrystalline surfaces is known to be favourable to the generation of the excess fluxes of self-interstitials and vacancies that exhibit the predominant crystallographic orientation along a <111> and <100> axis, respectively (Fig. 1a) (Bagraev et al. 2002; Bagraev et al. 2004a; Bagraev et al. 2004b; Bagraev et al. 2005). In the initial stage of the oxidation, thin oxide overlayer produces excess self-interstitials that are able to create small microdefects, whereas oppositely directed fluxes of vacancies give rise to their annihilation (Figs. 1a and 1b). Since the points of outgoing self-interstitials and incoming vacancies appear to be defined by the positive and negative charge states of the reconstructed silicon dangling bond (Bagraev et al. 2004a; Robertson 1983), the dimensions of small microdefects of the self-interstitials type near the Si (100) surface have to be restricted to 2 nm. Therefore, the distribution of the microdefects created at the initial stage of the oxidation seems to represent the fractal of the Sierpinski Gasket type with the built-in self-assembled Si-QW (Fig. 1b) (Bagraev et al. 2004a; Bagraev et al. 2004b; Bagraev et al. 2005). Then, the fractal distribution has to be reproduced by increasing the time of the oxidation process, with the $P_b$ centers as the germs for the next generation of the microdefects (Fig. 1c) (Robertson 1983; Gerardi et al. 1986). The formation of thick oxide overlayer under prolonged oxidation results in however the predominant generation of vacancies by the oxidized surface, and thus, in increased decay of these microdefects, which is accompanied by the self-assembly of the lateral silicon quantum wells (Fig. 1d).

Although Si-QWs embedded in the fractal system of self-assembled microdefects are of interest to be used as a basis of optically and electrically active microcavities in optoelectronics and nanoelectronics, the presence of dangling bonds at the interfaces prevents such an application. Therefore, subsequent short-time diffusion of boron would be appropriate for the passivation of silicon vacancies that create the dangling bonds during previous oxidation of the Si (100) surface thereby assisting the transformation of the arrays of microdefects in the neutral δ - barriers confining the ultra-narrow, 2nm, Si-QW (Fig. 1e, f and g).

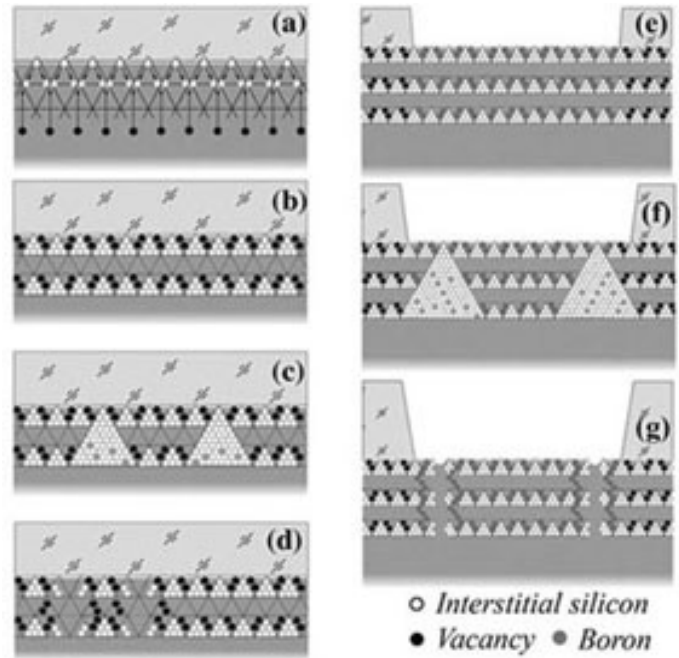

Fig. 1. A scheme of self-assembled silicon quantum wells (Si-QWs) obtained by varying the thickness of the oxide overlayer prepared on the Si(100) wafer. The white and black balls label the self-interstitials and vacancies forming the excess fluxes oriented crystallographically along a <111> and <100> axis that are transformed to small microdefects (a, b). The longitudinal Si-QWs between the alloys of microdefects are produced by performing thin oxide overlayer (b), whereas growing thick oxide overlayer results in the formation of additional lateral Si-QWs (d). Besides, medium and thick oxide overlayers give rise to the self-assembled microdefects of the fractal type (c). The atoms of boron replace the positions of vacancies in the process of subsequent short-time diffusion after making a mask and etching thereby passivating the alloys of microdefects and forming the neutral δ barriers that confine both the longitudinal (e, f) and lateral (g) Si-QWs.

We have prepared the p-type self-assembled Si-QWs with different density of holes ($10^9 \div 10^{12}$ cm$^{-2}$) on the Si (100) wafers of the n-type in frameworks of the conception discussed above and identified the properties of the two-dimensional high mobility gas of holes by the cyclotron resonance (CR), electron spin resonance (ESR), scanning tunneling spectroscopy (STM) and infrared Fourier spectroscopy techniques.

Firstly, the 0.35 mm thick n- type Si(100) wafers with resistivity 20 Ohm·cm were previously oxidized at 1150°C in dry oxygen containing $CCl_4$ vapors. The thickness of the oxide overlayer is dependent on the duration of the oxidation process that was varied from 20 min up to 24 hours. Then, the Hall geometry windows were cut in the oxide overlayer after preparing a mask and performing the subsequent photolithography. Secondly, the short-time diffusion of boron was done into windows from gas phase during five minutes at the diffusion temperature of 900°C. Additional replenishment with dry oxygen and the Cl levels into the gas phase during the diffusion process provided the fine surface injection of self-interstitials and vacancies to result in parity of the kick-out and vacancy-related diffusion mechanism. The variable parameters of the diffusion experiment were the oxide overlayer thickness and the Cl levels in the gas phase during the diffusion process (Bagraev et al. 2004a). The SIMS measurements were performed to define the concentration of boron, $5 \cdot 10^{21}$ cm$^{-3}$, inside the boron doped diffusion profile and its depth that was equal to 8 nm in the presence of thin oxide overlayer.

The Si-QWs confined by the δ - barriers heavily doped with boron inside the B doped diffusion profile were identified by the four-point probe method using layer-by-layer etching and by the cyclotron resonance (CR) angular dependencies (Figs. 2a and b). These CR measurements were performed at 3.8 K with a standard Brucker-Physik AG ESR spectrometer at X-band (9.1-9.5 GHz) (Bagraev et al. 1995; Gehlhoff et al. 1995). The rotation of the magnetic field in a plane normal to the diffusion profile plane has revealed the anisotropy of both the electron and hole effective masses in silicon bulk and Landau levels scheme in Si-QWs. This CR quenching and the line shifts for which a characteristic 180° symmetry was observed can be explained with the effect of the electrical field created by the confining potential inside p$^+$-diffusion profile and its different arrangement in longitudinal and lateral Si-QWs formed naturally between the δ barriers heavily doped with boron (Figs. 2a and b). The observed different behavior of the heavy and light holes may be explained by lifting the degeneracy between the $J_z = \pm 3/2$ and $J_z = \pm 1/2$ valence bands for $k = 0$ due to the confining potential.

The energy positions of two-dimensional subbands for the light and heavy holes in the Si-QW studied were determined by studying the far-infrared electroluminescence spectra obtained with the infrared Fourier spectrometer IFS-115 Brucker Physik AG (Fig. 3a) as well as by using the local tunneling spectroscopy technique (Bagraev et al. 2006b; Bagraev et al. 2007). The results obtained are in a good agreement with corresponding calculations following by Ref (Kotthaus and Ranvaud 1977; Weisbuch and Vinter 1991) if the width of the Si-QW, 2nm, is taken into account (Fig. 3b).

The STM technique was used to control the formation of the fractal distribution of the self-interstitials microdefects in the windows before and after diffusion of boron (Figs 4a). The self-assembled layers of microdefects inside the δ barriers that confine the Si-QW appear to be revealed by the STM method as the deformed potential fluctuations (DPF) after etching the oxide overlayer and after subsequent short-time diffusion of boron. The DPF effect induced by the microdefects of the self-interstitials type that are displayed as light poles in Figs. 4a is find to be brought about by the previous oxidation and to be enhanced by subsequent boron diffusion (Bagraev et al. 2000; Bagraev et al. 2004a). The STM images demonstrate that the ratio between the dimensions of the microdefects produced during the different stages of the oxidation process is supported to be equal to 3.3 thereby defining the self-assembly of microdefects as the self-organization of the fractal type (Figs. 4b and 1f). The analysis of the STM image in detail has shown that the dimension of the smallest microdefect observed in fractal series, ~2nm, is consistent with the parameters expected from the tetrahedral model of the $Si_{60}$ cluster (Fig. 4c) (Bao-xing Li et al. 2000).

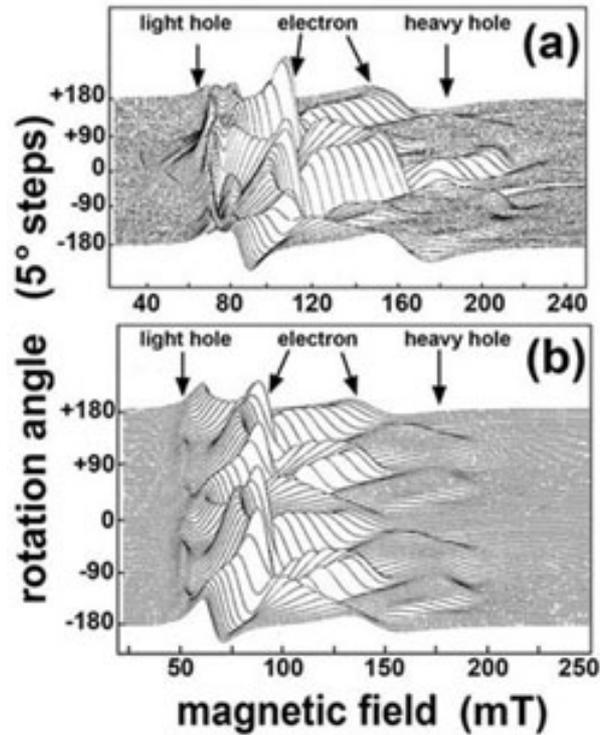

Fig. 2. Cyclotron resonance spectra for the ultra-shallow boron diffusion profiles obtained on the n - type silicon {100} surfaces at the diffusion temperatures of 900°C (*a*) and 1100°C (*b*) which consist of the δ barriers confining the longitudinal (a) and lateral (b) Si-QW. Rotation of magnetic field *B* in a {110}-plane perpendicular to a {100}-surface of profiles (0° = $B \perp$ surface; ± 90° = $B \parallel$ surface), *T*= 3.8 K, *ν* = 9.45 GHz.

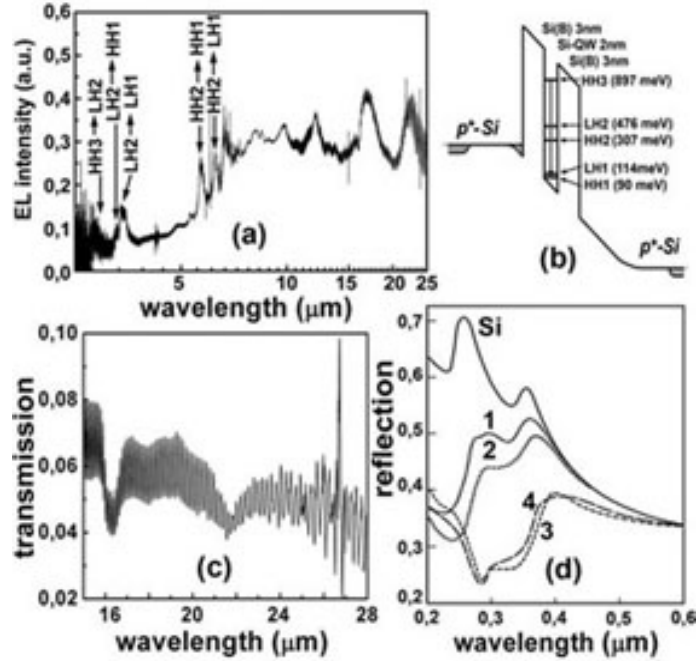

Fig. 3. Electroluminescence spectrum (a) that defines the energies of two-dimensional subbands of heavy and light holes in the p-type Si-QW confined by the δ - barriers heavily doped with boron on the n-type Si (100) surface (b).
(c) Transmission spectrum that reveals both the local phonon mode, $\lambda$ = 16.4 μm, and the superconductor gap, $\lambda$ = 26.9 μm, manifestation. (d) The reflection spectra from the n - type Si (100) surface and from the ultra-shallow boron diffusion profiles prepared on the n - type Si (100) surface that consist of the δ barriers confining the ultra-narrow Si-QW. The curves 1-4 are related to the δ barriers with different concentration of boron. The values of the concentration boron in different samples are characterized by the following ratio: curve 1 – 0.2, 2 – 0.3, 3 – 0.35, 4 -0.4. The concentration of boron in the sample characterized by the fourth curve is equal to $5 \cdot 10^{21}$ cm$^{-3}$.

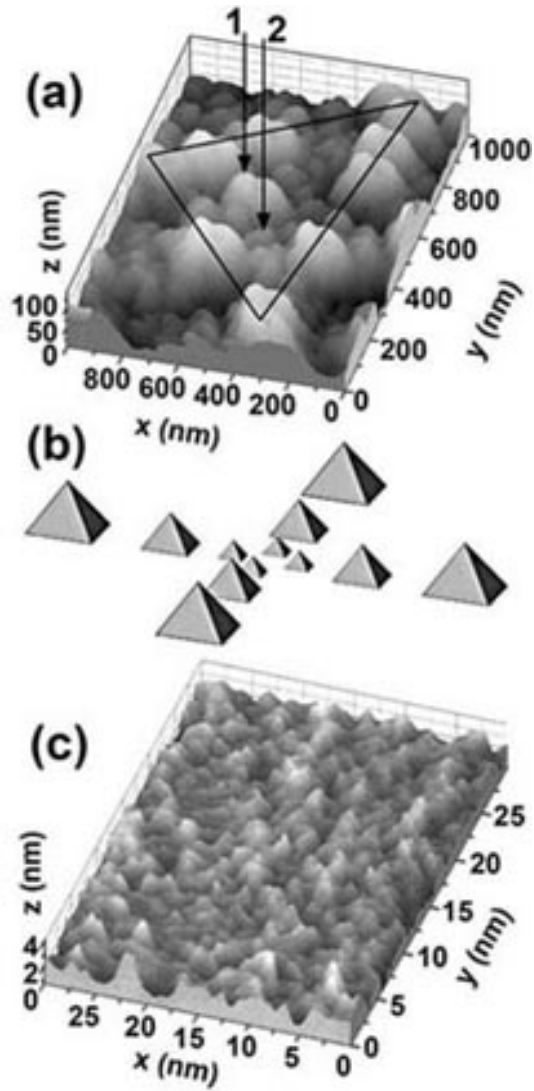

Fig. 4. (a) - STM image of the ultra-shallow boron diffusion profile prepared at the diffusion temperature of 800°C into the Si(100) wafer covered previously by medium oxide overlayer X∥[001], Y∥[010], Z∥[100]. Solid triangle and arrows that are labeled as 1 and 2 exhibit the microdefects with dimensions 740 nm, 225 nm and 68 nm, respectively, which are evidence of their fractal self-assembly.

(b) - The model of the self-assembled microcavity system formed by the microdefects of the fractal type on the Si(100) surface.

(c) - STM image of the ultra-shallow boron diffusion profile prepared at diffusion temperature of 900°C into the Si(100) wafer covered previously by medium oxide overlayer. X∥[001], Y∥[010], Z∥[100].

Thus, the δ - barriers, 3 nm, heavily doped with boron, 5 $10^{21}$ cm$^{-3}$, represent really alternating arrays of the smallest undoped microdefects and doped dots with dimensions restricted to 2 nm (Fig. 4c). The value of the boron concentration determined by the SIMS method seems to indicate that each doped dot located between undoped microdefects contains two impurity atoms of boron. Since the boron dopants form shallow acceptor centers in the silicon lattice, such high concentration has to cause a metallic-like conductivity. Nevertheless, the angular dependencies of the cyclotron resonance spectra demonstrate that the p-type Si-QW confined by the δ - barriers heavily doped with boron contains the high mobility 2D hole gas which is characterized by long transport relaxation time of heavy and light holes at 3.8 K, $\tau \geq 5 \cdot 10^{-10}$ s (Figs. 2a and b) (Bagraev et al. 1995; Gehlhoff et al. 1995; Bagraev et al. 2005). Thus, the transport relaxation time of holes in the ultra-narrow SQW appeared to be longer than in the best MOS structures contrary to what might be expected from strong scattering by the heavily doped δ - barriers. This passive role of the δ - barriers between which the Si-QW is formed was quite surprising, when one takes into account the high level of their boron doping. To eliminate this contradiction, the ESR technique has been applied for the studies of the boron centers packed up in dots (Bagraev et al. 2002; Bagraev et al. 2005).

The angular dependences of the ESR spectra at different temperatures in the range 3.8÷27 K that reveal the trigonal symmetry of the boron dipole centers have been obtained with the same ESR spectrometer, the Brucker-Physik AG ESR spectrometer at X-band (9.1-9.5 GHz), with the rotation of the magnetic field in the {110}-plane perpendicular to a {100}-interface ($B_{ext}$ = 0°, 180° parallel to the Si-QW plane, $B_{ext}$ = 90° perpendicular to the Si-QW plane) (Figs. 5a, b, c and d). No ESR signals in the X-band are observed, if the Si-QW confined by the δ - barriers is cooled down in the external magnetic field ($B_{ext}$) weaker than 0.22 T, with the persistence of the amplitude and the resonance field of the trigonal ESR spectrum as function of the crystallographic orientation and the magnetic field value during cooling down process at $B_{ext} \geq 0.22$ T (Figs. 5a, b and c). With increasing temperature, the ESR line observed changes its magnetic resonance field position and disappears at 27 K (Fig. 5c). The observation of the ESR spectrum is evidence of the fall in the electrical activity of shallow boron acceptors contrary to high level of boron doping. Therefore, the trigonal ESR spectrum observed seems to be evidence of the dynamic magnetic moment that is induced by the exchange interaction between the small hole bipolarons which are formed by the negative-U reconstruction of the shallow boron acceptors, $2B^0 \rightarrow B^+ + B^-$, along the <111> crystallographic axis (Fig. 6a) (Slaoui et al. 1983; Gehlhoff et al. 1995; Bagraev et al. 2002). These small hole bipolarons localized at the dipole boron centers, $B^+$ - $B^-$, seem to undergo the singlet-triplet transition in the process of the exchange interaction through the holes in the Si-QW thereby leading to the trigonal ESR spectrum (Fig. 5a, b, c and d). Besides, the sublattice of the hole bipolarons located between the undoped microdefects appears to define the one-electron band scheme of the δ - barriers as well as the transport properties for the 2D gas of holes in the Si-QW (Figs. 6b and 3b) (Bagraev et al. 2002).

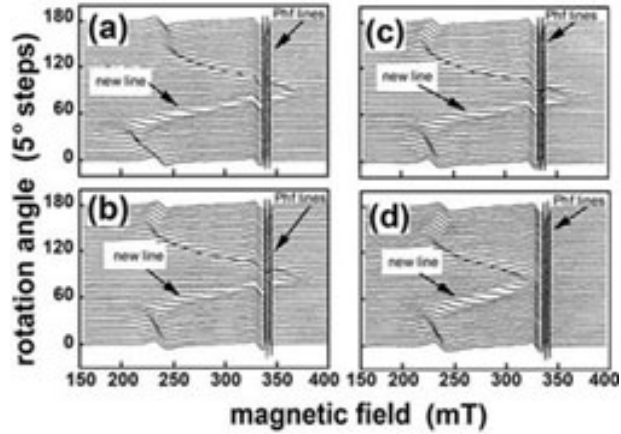

Fig. 5. The trigonal ESR spectrum observed in field cooled ultra-shallow boron diffusion profile that seems to be evidence of the dynamic magnetic moment due to the trigonal dipole centers of boron inside the δ - barriers confining the Si-QW which is persisted by varying both the temperature and magnetic field values. $B_{ext}$ ∥ <110> (a), ∥ <112> (b), ∥ <111> (c, d). Rotation of the magnetic field in the {110}-plane perpendicular to a {100}-interface ($B_{ext} = 0°, 180°$ ∥ interface, $B_{ext} = 90°$ ⊥ interface), ν = 9.45 GHz, T = 14 K (a, b, c) and T=21 K (d).

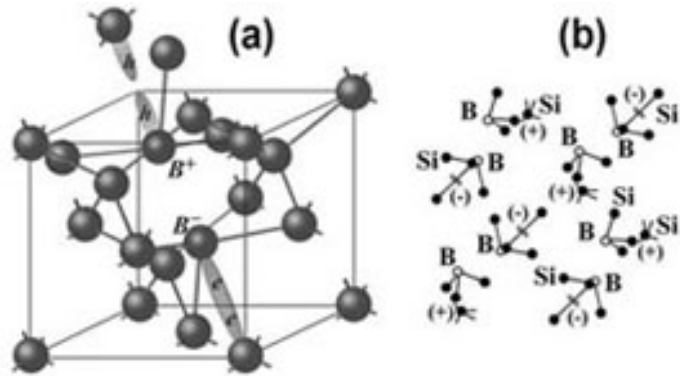

Fig. 6. (a) Model for the elastic reconstruction of a shallow boron acceptor which is accompanied by the formation of the $C_{3V}$ dipole ($B^+ - B^-$) centers as a result of the negative-U reaction: $2B^o \rightarrow B^+ + B^-$. (b) A series of the dipole negative-U centers of boron located between the undoped microdefects that seems to be a basis of nanostructured δ - barriers confining the Si-QW.

In order to determine the one-electron band scheme of the δ - barriers that confine the Si-QW, the reflection spectra $R(\lambda)$ were studied using a UV-VIS Specord M-40 spectrophotometer with an Ulbricht sphere for the reflectivity measurements (Bagraev et al. 2000). Fig. 3d shows the spectra of the reflection from the δ - barriers with different concentration of boron. The decrease in $R(\lambda)$ compared with the data of the silicon single crystal and the drops in the position of the peaks at the wavelengths of $\lambda$=354 and 275 nm are observed. The above peaks are related to the transitions between Γ-L valleys and in the vicinity of the point X in the Brillouin zone, with the former of the above peaks being assigned to the direct transition $\Gamma'_{25} - \Gamma'_2$, whereas the latter peak is attributed to the transition $X_4 - X_1$ (Slaoui et al. 1983). An analysis of the spectral dependence of the reflection coefficient shows that the presence of the microcavities formed by the self-assembled microdefects with medium size reduces $R(\lambda)$ most profoundly in the short-wavelength region of the spectrum (200-300 nm). It follows from the comparison of $R(\lambda)$ with the STM data that the position of the minima in the reflection coefficient in the spectral dependence $R(\lambda)$ and the microcavity size are interrelated and satisfy the Bragg condition, $x = \lambda/2n$, where $x$ is the cavity size, $\lambda$ is the wavelength, and $n$ is the refractive index of silicon, $n$=3.4 (see Fig. 4a). The $R(\lambda)$ drop in the position of the $\Gamma'_{25} - \Gamma'_2$ and $X_4 - X_1$ transitions appears to be due to the formation of the wide-gap semiconductor layer with increasing the concentration of boron. These data substantiate the assumption noticed above that the dot containing the small hole bipolaron is to establish the band structure of the δ - barrier with the energy confinement more than 1.25eV in both the conduction and the valence band of the Si-QW (Fig. 3d).

### 3. Superconductor properties for δ - barriers heavily doped with boron

In common with the other solids that contain small onsite localized small bipolarons (Anderson 1975; Watkins 1984; Street et al. 1975; Kastner et al. 1976; Baraff et al. 1980; Bagraev and Mashkov 1984; Bagraev and Mashkov 1988), the δ - barriers containing the dipole boron centres have been found to be in an excitonic insulator regime at the sheet density of holes in the Si-QW lower than $10^{15}$ m$^{-2}$. The conductance of these silicon nanostructures appeared to be determined by the parameters of the 2D gas of holes in the Si-QW (Bagraev et al. 2002; Bagraev et al. 2004b; Bagraev et al. 2006a). However, here we demonstrate using the electrical resistance, thermo-emf, specific heat magnetic susceptibility and local tunnelling spectroscopy techniques that the high sheet density of holes in the Si-QW ($>10^{15}$ m$^{-2}$) gives rise to the superconductor properties for the δ - barriers which result from the transfer of the small hole bipolarons through the negative-U centers (Šimánek 1979; Ting et al. 1980; Alexandrov and Ranninger 1981; Chakraverty 1981; Alexandrov and Mott 1994) in the interplay with the multiple An-dreev reflections inside the Si-QW (Andreev 1964; Klapwijk 2004; van Dam et al. 2006; Jarillo-Herrero et al. 2006; Jie Xiang et al. 2006).

The resistance, thermo-emf and Hall measurements of the device with high density of 2D holes, $6 \cdot 10^{15}$ m$^{-2}$, performed within Hall geometry were made in Special Design Electric and Magnetic Measurement System with high precision bridge (Fig. 7a). The identical device was used in the studies of the local tunneling spectroscopy with the STM spectrometer to register the tunneling current as a function of the voltage applied between the STM tip and the Hall contacts (Fig. 7b). The measurements in the range 0.4-4 K and 1.2-300 K were carried out respectively in a He$^3$ and He$^4$ cryostat.

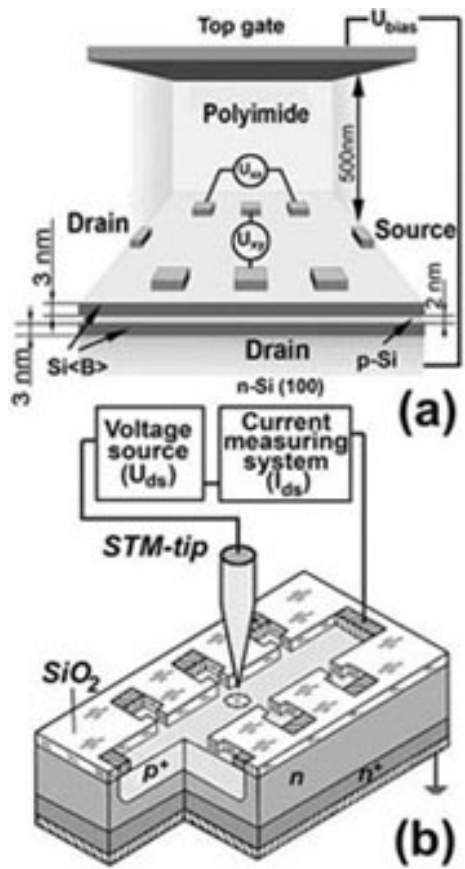

Fig. 7. (a) Schematic diagram of the devices that demonstrates a perspective view of the p-type Si-QW confined by the δ - barriers heavily doped with boron on the n-type Si (100) surface. The top gate is able to control the sheet density of holes and the Rashba SOI value. The depletion regions indicate the Hall geometry of leads.
(b) Planar field-effect silicon transistor structure with the STM tip, which is based on an ultra-shallow p$^+$-diffusion profile prepared in the Hall geometry. The circle dashed line exhibits the point STM contact region.

The current-voltage characteristics (CV) measured at different temperatures exhibited an ohmic character, whereas the temperature dependence of the resistance of the device is related to two-dimensional metal only in the range 220-300 K (Fig. 8a). Below 220 K the resistance increases up to the value of 6.453 kOhm and then drops reaching the negligible value at the temperature of 145 K. The creation of the additional peak when the resistance begins to fall down seems to be evidence of the superconductor properties caused by the transfer of the small hole bipolarons. This peak shows the logarithmic temperature dependence that appears to be due to the Kondo-liked scattering of the single 2D holes tunneling through the negative-U boron dipole centres of boron at the Si-QW – δ-barrier interfaces.

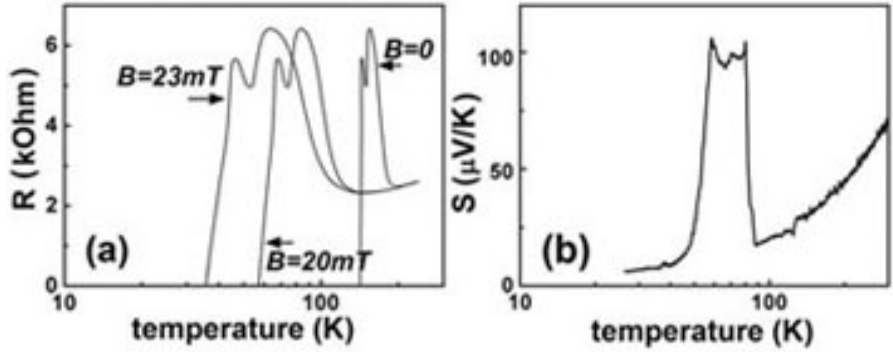

Fig. 8. The resistance (a) and thermo-emf (Seebeck coefficient) (b) temperature dependences that were observed in the ultra-shallow p$^+$-diffusion profile which contains the p-type Si-QW confined by the δ-barriers heavily doped with boron on the n-type Si (100) surface.

As was to be expected, the application of external magnetic field results in the shift of the resistance drop to lower temperatures, which is accompanied by the weak broadening of the transition and the conservation of the peak values of the resistance (Fig. 8a). Since similar peaks followed by the drops of the Seebeck coefficient value are revealed also in the temperature dependences of the thermo-emf (Fig. 8b), the Kondo-liked scattering seems to be the precursor of the optimal tunneling of single holes into the negative-U boron centers of boron (Trovarelli et al. 1997). This process is related to the conduction electron tunneling into the negative-U centers that is favourable to the increase of the superconducting transition temperature, $T_c$, in metal-silicon eutectic alloys (Šimánek 1979; Ting et al. 1980). The effect of single-hole tunneling is also possible to resolve some bottlenecks in the bipolaronic mechanism of the high temperature superconductivity, which results from the distance between the negative-U centers lesser than the coherence length (Alexandrov and Ranninger 1981; Alexandrov and Mott 1994). Besides, two experimental facts are needed to be noticed. Firstly, the maximum value of the resistance, 6.453 kOhm ≈ $h/4e^2$, is independent of the external magnetic field. Secondly, applying a magnetic field is surprisingly to stabilize the δ-barrier in the state of the two-dimensional metal up to the temperature value corresponding to the shift of a transition to lower temperatures (Fig. 8a). Thus, the δ-barriers confining Si-QW seem to be self-organized as graphene (Geim and Novoselov 2007) owing to heavily doping with boron which gives rise to the formation of the negative-U dipole centers.

The value of the critical temperature, $T_c$=145 K, the estimations of the superconductor gap, $2\Delta$=0.044 eV, and the $T$=0 upper critical field, $H_{C2}$=0.22 T, that were derived from the resistance and thermo-emf measurements using well-known relationships $2\Delta=3.52k_BT_c$ and $H_{c2}(0)=-0.69(dH_{c2}/dT|_{T_c})T_c$ (Werthamer et al. 1966) appear to be revealed also in the temperature and magnetic field dependencies of the static magnetic susceptibility obtained by the Faraday balance method (Fig. 9).

These dependences were measured in the range 3.5-300 K with the magnetic balance spectrometer MGD312FG. High sensitivity, $10^{-9} \div 10^{-10}$ CGS, should be noted to be provided by the $B\,dB/dx$ stability using this installation. Pure InP samples with the shape and size similar to the silicon samples studied here that are characterized by temperature stable magnetic susceptibility, $\chi = 313 \cdot 10^{-9}$ cm$^3$/g, were used to calibrate the $B\,dB/dx$ values.

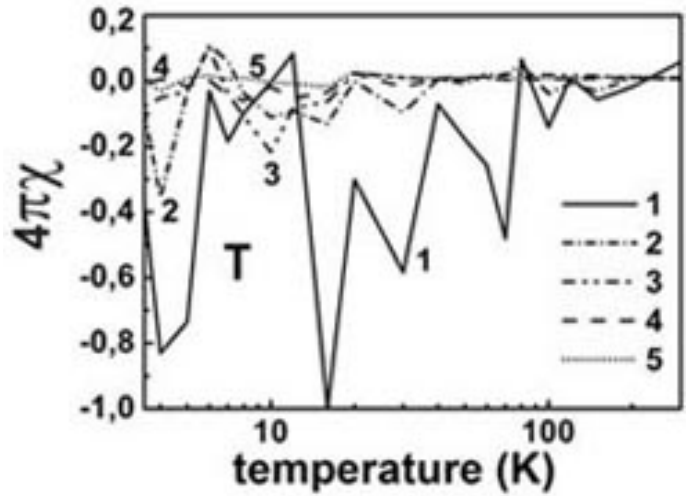

Fig. 9. Static magnetic susceptibility vs temperature that was observed in field-cooled ultra-shallow p$^+$-diffusion profile which contains the p-type Si-QW confined by δ-barriers heavily doped with boron on the n-type Si (100) surface. Diamagnetic response is revealed by field-out procedure: 1 - 10 mT; 2 – 20 mT, 3 – 30 mT; 4 – 40 mT; 5 – 50 mT.

The value of temperatures corresponding to the drops of the diamagnetic response on cooling is of importance to coincide with the drops of the resistance and the Seebeck coefficient thereby confirming the role of the charge correlations localized at the negative-U dipole centers in the Kondo-liked scattering and the enhancement of the critical temperature (Fig. 9). Just the same temperature dependence of the paramagnetic response observed after the field-in procedure exhibits the effect of the arrays of the Josephson transitions revealed by the STM image (Fig. 4c) on the flux pinning processes in the superconductor δ-barriers heavily doped with boron (Bagraev et al. 2006b). The plots of the magnetic susceptibility vs temperature and magnetic field shown in Figs. 10a result in the value of $H_{C2}$, $H_{C2}$=0.22 T, that corresponds to the data obtained by the measurements of the resistance and allow the estimation of the coherence length , ξ=39 nm, where ξ = $(\Phi_0/2\pi H_{C2})^{1/2}$, $\Phi_0$ =h/2e. This value of the coherence length appears to be in a good agreement with the estimations of the superconductor gap, 2Δ=0.044 eV, made if the value of the critical temperature, $T_C$=145 K, is taken into account, $\xi = 0.18\hbar v_F / k_B T_c$, where $v_F$ is the Fermi velocity, and with the first critical magnetic field, $H_{C1}$=215 Oe, defined visually from Fig. 10a.

The oscillations of the magnetic susceptibility value revealed by varying both the temperature and magnetic field value seem to be due to the vortex manipulation in nanostructured δ-barriers (Figs. 10b and c). Since the fractal series of silicon microdefects identified by the STM images is embedded in the superconductor δ-barrier, the multi-quanta vortex lattices are able to be self-organized (Vodolazov et al. 2007). These self-assembled pinning arrays that can be simulated as a series of anti-dots appear to capture in consecutive order several vortices and thus to enhance critical current (de Souza Silva et al. 2006; Vodolazov et al. 2007). Furthermore, the upper critical field, $H_{C2}$, is evidently dependent step-like on both temperature and magnetic field, because the critical current increases jump-like each time when the regular vortex is captured at such an anti-dot that is revealed by the corresponding oscillations of the diamagnetic response (Fig. 10c). The period of these oscillations that is derived from the plots in Fig. 10c appears to be due to the distance between the small microdefects in the fractal series identified by the STM image, ≈ 120 nm, with average dimensions equal to 68 nm (Fig. 4a): ΔB·S=$\Phi_0$, where ΔB is the period oscillations, S = π$d^2$/4, d is the distance between anti-dots (≈ 120 nm). The dependence $H_{C2}(T)$ is of importance to be in a good agreement with the value of this period, because each maximum of the diamagnetic response as a function of magnetic field is accompanied by the temperature satellite shifted by approximately 140 K (~$T_C$) to higher temperatures. In addition to the oscillations of the magnetic susceptibility, the B-T diagram shown in Fig. 10b exhibits also the quantization of the critical current which seems to be caused by the vortex ratchet effect (de Souza Silva et al. 2006).

The enhancement of the critical current due to the N $\Phi_0$ vortex capture at the anti-dots seems to result also from the studies of a specific heat anomaly at $T_C$ (Figs. 11a and b). This anomaly arises at the temperature of 152 K (H=0) that is close to the value of the critical temperature derived from the measurements of the resistance and the magnetic susceptibility. With increasing external magnetic field, the position of the jump in specific heat is shifted to the range of low temperatures (Fig. 11a). The jump values in specific heat, $\Delta C$, appear to be large if the abnormal small effective mass of heavy holes in these 'sandwich' structures, S-Si-QW-S, is taken into account to be analyzed within frameworks of a weak coupled BCS superconductor (Bagraev et al. 2008a). The oscillations of a specific heat anomaly as a function of external magnetic field are seen to be in a good agreement with the corresponding behavior of the diamagnetic response that corroborates additionally the important role of vortices in the superconductor properties of the nanostructured δ-barriers (Fig. 11b).

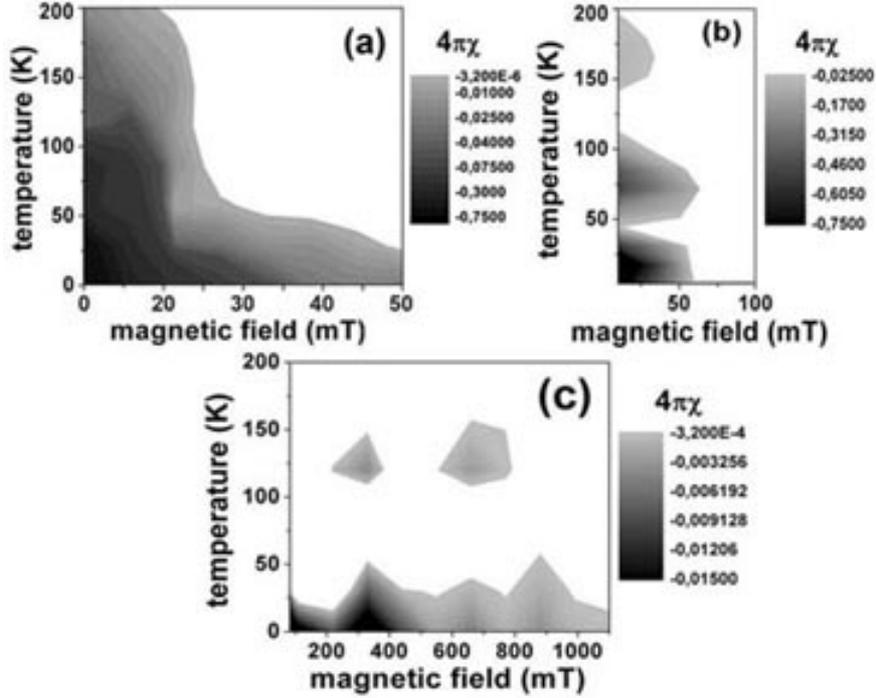

Fig. 10. Plots of static magnetic susceptibility vs temperature and magnetic field that was observed in field-cooled ultra-shallow $p^+$-diffusion profile which contains the p-type Si-QW confined by δ-barriers heavily doped with boron on the n-type Si (100) surface. Diamagnetic response (a) revealed by field-out procedure demonstrates also the oscillations that seem to be related to the ratchet effect (b) and the quantization of the critical current (c).

The values of the superconductor energy gap derived from the measurements of the critical temperature using the different techniques appear to be practically identical, 0.044 eV. Nevertheless, the direct methods based on the principles of the tunneling spectroscopy are necessary to be applied for the identification of the superconductor gap in the δ-barriers confining the Si-QW (Figs. 7a and b). Since the nanostructured δ-barriers are self-assembled as the dots containing a single dipole boron center that alternate with undoped silicon anti-dots shown in Fig. 4c, the tunneling current can be recorded by the applied voltage to the contacts prepared in the Hall geometry (Fig. 7a). The tunneling current-voltage characteristic obtained is direct evidence of the superconductor gap that appears to be equal to 0.044 eV (Fig. 12a) (Bagraev et al. 1998). To increase the resolution of this experiment, a series of doped dots - undoped anti-dots involved in the sequence measured should not possess large discrepancies in the values of the superconductor energy gap. Therefore, the one-dimensional constriction is expediently to be prepared for the precise measurements of the tunneling current-voltage characteristics (Bagraev et al. 2002; Bagraev et al. 2004b; Bagraev et al. 2005; Bagraev et al. 2006a).

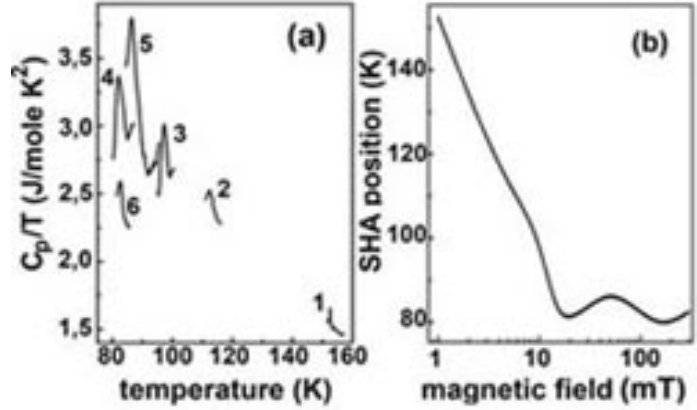

Fig. 11. (a) Specific heat anomaly as C/T vs T that seems to reveal the superconducting transition in field-cooled ultra-shallow $p^+$-diffusion profile which contains the p-type Si-QW confined by δ-barriers heavily doped with boron on the n-type Si (100) surface. Magnetic field value: 1- 0 mT; 2 – 5 mT, 3 – 10 mT; 4 – 21.5 mT; 5 – 50 mT; 6 – 300 mT. (b) The oscillations of a specific heat anomaly as a function of external magnetic field that seem to be due to the quantization of the critical current.

The other way for the definition of the superconductor energy gap is to use the techniques of the local tunneling spectroscopy (LTS) (Suderow et al. 2002; Bagraev et al. 2005; Fischer et al. 2007). The local density of states (LDOS) can be accessed by measuring the tunnelling current, while the bias voltage is swept with the tip held at a fixed vertical position (Fig. 7b) (Fischer et al. 2007). If a negative bias voltage is applied to the δ-barriers, holes will tunnel into unoccupied sample states, whereas at a positive bias voltage they will tunnel out of occupied sample states. Since the transport conditions inside the 'sandwich' structures, S-Si-QW-S, are close to ideal (Bagraev et al. 2002; Bagraev et al. 2004b; Bagraev et al. 2005; Bagraev et al. 2006a), the tunnelling conductance, *dI/dV(V)*, provides the measurements of the LDOS thereby allowing the precise definition of the superconductor energy gap. The LTS current-voltage characteristic shown in Fig. 12b that has been registered in the studies of the device structure identical discussed above demonstrates also the value of the superconductor energy gap equal to 0.044 eV which is in self-agreement with the measurements of the critical temperature and the upper critical magnetic field.

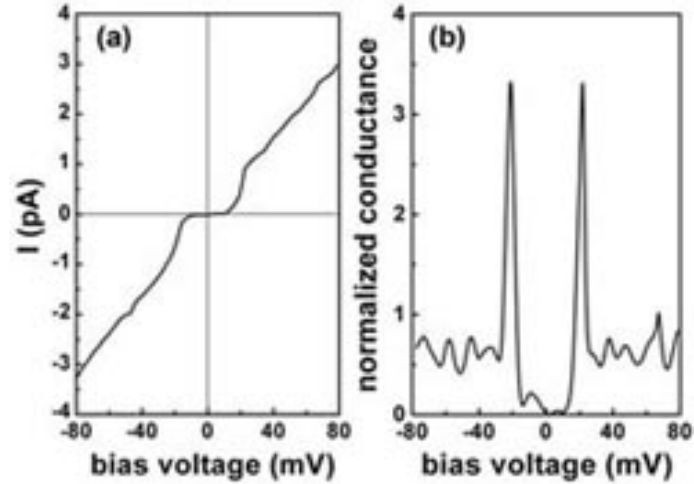

Fig. 12. The *I-U* (a) and *dI/dV(V)* (b) characteristics found by the current-voltage measurements (a) and using the STM point contact technique (b), which identify the superconductor energy gap in the nanostructured δ-barriers heavily doped with boron that confine the Si-QW on the n-type Si (100) surface. (a) – 77 K; (b) – 4.2 K.

In order to identify the transfer of the small hole bipolarons as a possible mechanism of nano-supeconductivity, the transport of holes in the S-Si-QW-S structures is followed to be studied at different orientation of the external magnetic field relatively the Si-QW plane. The dependences of the longitudinal and Hall voltages on the magnetic field value shown in Figs. 13a, b and c are evidence of the Zeeman effect that seems to be due to the creation of the triplet and singlet states of the small hole bipolarons localized at the dipole boron centers (Fig. 6a). The sign inversion of both $U_{xx}$ and $U_{xy}$ voltages is of importance to result from the change of the magnetic field direction to opposite. Thus, the transport of the small hole (bi)polarons that are able to capture and/or scattered on the dipole boron centers seems to be caused by the diamagnetic response induced by applying a magnetic field.

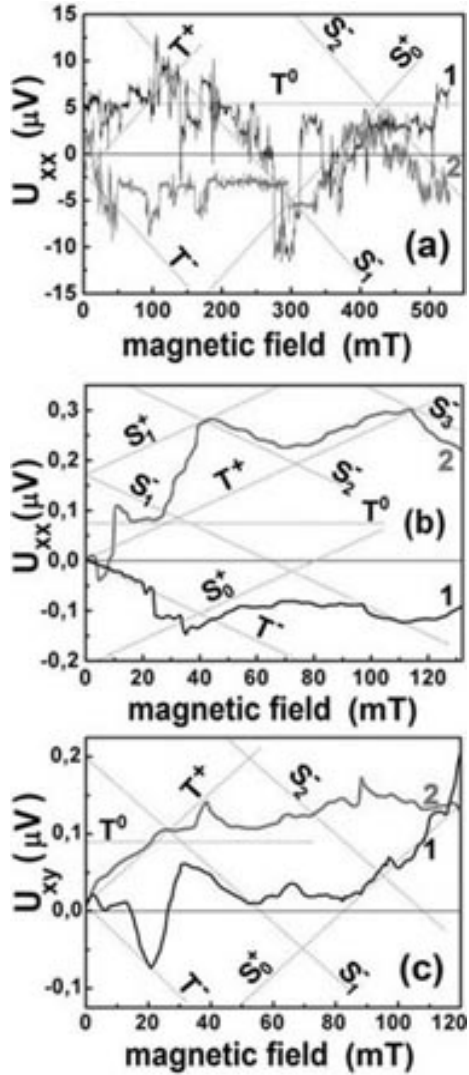

Fig. 13. (a) $U_{xx}$ vs the value of the magnetic field applied perpendicularly to the plane of the p-type Si-QW confined by the δ-barriers on the n-type Si (100) surface. $I_{ds}$=10 nA. T=77 K. Curves 1 and 2 measured for two orientations of a magnetic field reveal the sign of $U_{xx}$ that corresponds to the diamagnetic response of the superconductor δ-barriers.
$U_{xx}$ (b) and $U_{xy}$ (c) vs the value of the magnetic field applied in the plane of the p-type Si-QW confined by the δ-barriers on the n-type Si (100) surface, B ∥ XY. $I_{ds}$=10 nA. T=77 K. (c) Curves 1 and 2 measured for two orientations of a magnetic field along XY are evidence of the diamagnetic response of the superconductor δ-barriers, because the sign of $U_{xy}$ results from the circulating current that is oppositely to the direction of a magnetic field.

Besides, the magnetic field dependences of the $U_{xx}$ and $U_{xy}$ voltages considered within frameworks of the triplet, $T^+$, $T^0$, $T^-$, as well as the ground, $S_0^+$, $S_0^-$, and excited, $S_1^+$, $S_1^-$, states undergone by the Zeeman splitting appear to reveal the presence of the upper critical magnetic field $H_{c2}$ and the oscillations of the critical current which are in a good agreement with the measurements of the magnetic susceptibility (see Fig. 13a and Figs. 10a, b, c). The resonance behaviour of the $U_{xx}(H)$ and $U_{xy}(H)$ dependences in the anti-crossing points of the triplet sublevels ($T^+$-$T^0$) is evidence of the spin polarization that results from the selective population or depopulation of the $T^+$ and $T^-$ states relatively to the $T^0$ state in consequence of the partial removal of a ban on the forbidden triplet-singlet transitions (Laiho et al. 1998). The spin polarization of the (bi)polarons in the triplet state in the S-Si-QW-S structures should be of importance in the studies of the spin interference caused by the Rashba spin-orbit interaction in the quantum wires and rings (Bagraev et al. 2006a; Bagraev et al. 2008a). The creation of the excited singlet states in the processes of the bipolaronic transport is also bound to be noticed, because owing to the transitions from the excited to the ground singlet state of the small hole bipolarons these 'sandwich' structures seem to be perspective as the sources and recorders of the GHz emission that is revealed specifically in the electroluminescence spectra as a low frequency modulation (see Fig. 3a). The optical detection of magnetic resonance of the single impurity centers in the Si-QW confined by the δ-barriers heavily doped with boron was espeshially performed by the direct measurements of the transmission spectra under such an internal GHz emission in the absence of the external cavity resonator (Bagraev et al. 2003a; Bagraev et al. 2003b).

The extraordinary properties of the p-type Si-QW confined by the δ - barriers heavily doped with boron emerge in the Shubnikov – de Haas (SdH) oscillations shown in Fig. 14a that were revealed by the measurements of the diamagnetic response at high temperatures, 77 K (curve 2, Fig. 13a). These findings became it possible owing to the small effective mass of the heavy holes that was controlled by studying the temperature dependences of the SdH oscillations with the combination of the CR data and the estimations from the period of the Aharonov-Casher (AC) oscillations in the low sheet density Si-QW (Bagraev et al. 1995; Gehlhoff et al. 1995; Bagraev et al. 2008a; Bagraev et al. 2008b). The high sensitivity of the diamagnetic response to the variations of a magnetic field appears to provide the observation of the SdH oscillations at ν < 1 in the relatively weak magnetic fields, where ν is the number of the Landau level.

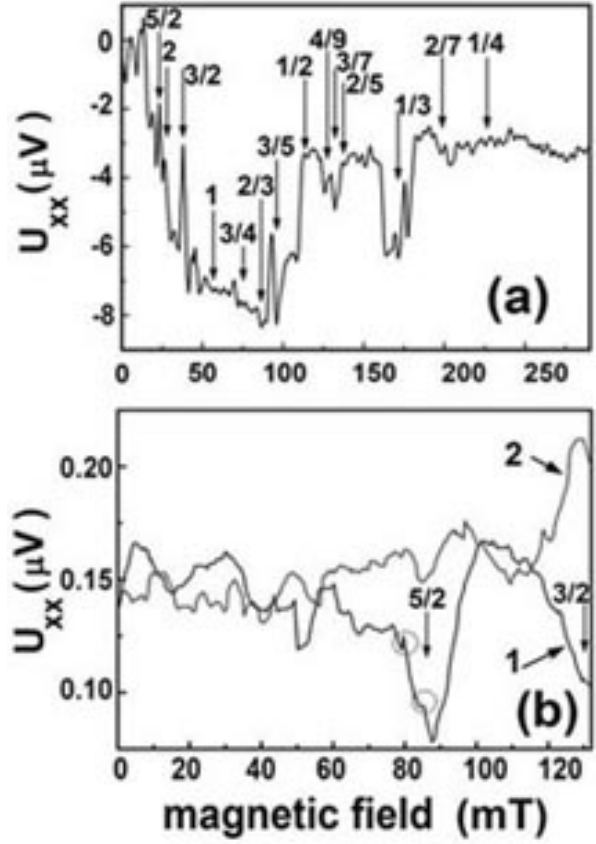

Fig. 14. Shubnikov - de Haas oscillations in the external magnetic field applied perpendicularly (a) and in plane (b) of the p-type Si-QW confined by the δ-barriers on the n-type Si (100) surface. $I_{ds}$=10 nA. T=77 K. (b) Curves 1 and 2 measured for two orientations of a magnetic field along XX are evidence of the diamagnetic response of the superconductor δ-barriers, because the sign of $U_{xx}$ results from the circulating current that is oppositely to the direction of a magnetic field. Dashed circles depict the closest AAS features.
(a) $p_{2D}$ = 3 $10^{13}$ m$^{-2}$; $m^*$ = 5.2 $10^{-4}$ $m_0$. (b) $p_{2D}$ = 1.1 $10^{14}$ m$^{-2}$; $m^*$ = 2.6 $10^{-4}$ $m_0$.

The SdH oscillations are very surprisingly to be observed in the external magnetic field that is in parallel with the Si-QW plane (Fig. 14b) along with the ordinary SdH (Fig. 14a). Since the emf, $U_{xx}$, of the opposite sign relatively to the direction of the longitudinal magnetic field, is observed, the SdH oscillations shown in Fig. 14b seem to result also from the diamagnetic response from the superconductor δ - barriers. This premise is supported by the presence of additional features superimposed on the oscillations which are caused by the Al'tshuler – Aronov - Spivak (AAS) oscillations taking account of the Si-QW width, $d$=2 nm, and the distance between the Hall contacts, $b$, equal to 0.2 mm (see dashed circles in Fig. 14b, where the distance between dashed circles, $\Delta B$, corresponds to the relationship $\Delta B \cdot S = \Phi_0/2$; $S=d \cdot b$) (Altshuler et al. 1981). The effective mass of holes in the 'sandwich' S-Si-QW-S structure seems to be estimated if the dependence of the amplitude of the $U_{xx}$ oscillations, $\Delta U_{xx}$, on the ν value crossing the Fermi level, is taken into account, $\nu \hbar \omega_c = \nu \hbar eB/m^* = e\Delta U_{xx} N$, where $N$ is the number of carriers involved in the circulation around a closed path along the Si-QW in opposite to the longitudinal magnetic field; the closed path corresponds to the section of the 'sandwich' structure. $N$ can be calculated from the value of the critical current revealed by the diamagnetic response and/or taking into account the quantum resistance of the δ - barrier, $h/4e^2$, before the transition to the superconductor state (Fig. 8a). Both versions of the $N$ definition appeared to be in a self-agreement that made it possible the findings of the $p_{2D}$ and $m^*$ values derived from Fig. 14b: $m^* = \nu \hbar B / N\Delta U_{xx}$. The values obtained, $p_{2D}$ = 1.1 $10^{14}$ m$^{-2}$; $m^*$ = 2.6 $10^{-4}$ $m_0$, agree satisfactorily with the data that result from the CR and AC measurements (Bagraev et al. 1995; Gehlhoff et al. 1995; Bagraev et al. 2008a; Bagraev et al. 2008b).

Notice that the studies of the SdH oscillations reveal a marked decrease in the sheet density of holes in the magnetic fields lower than the upper critical field, $H_{c2}$, as compared with the value of $p_{2D}$ derived from the Hall measurements in high magnetic fields, 6 $10^{15}$ m$^{-2}$. These variations from the maximum value of $p_{2D}$ in the Si-QW seem to be caused by outgoing the 2D holes in the δ - barriers during the superconductor transition.

Thus, the extremely low value of the hole effective mass in the 'sandwich' S-Si-QW-S structures seems to be the principal argument for the bipolaronic mechanism of high temperature superconductor properties that is based on the coherent tunneling of (bi)polarons (Alexandrov and Ranninger 1981; Alexandrov and Mott 1994). The local phonon mode manifestation at λ = 16.4 μm that presents, among the superconductor gap, λ = 26.9 μm ⇔ Δ= 0.046 eV, in the transmission spectrum favours the use of this conception (Fig. 3c). High frequency local phonon mode, 76 meV, appears to exist simultaneously with the intermediate value of the coupling constant, $\kappa$.

The value of the coupling constant, $\kappa = VN(0)$, is derived from the BCS formula $\Delta = 2\hbar \omega_D \exp(-1/\kappa)$ taking account of the experimental values of the su-perconductor energy gap, $2\Delta = 0.044$ eV, and the local phonon mode energy, $\hbar\omega_D = 76$ meV. This estimation results in $\kappa \approx 0.52$ that is outside the range 0.1÷0.3 for metallic low-temperature superconductors with weak coupling described within the BCS approach. Therefore the superconductor properties of the 'sandwich' S-Si-QW-S structures seem to be due to the transfer of the mobile small hole bipolarons that gives rise to the high $T_c$ value owing to small effective mass. In support of this interpretation it should be noted that the superconductivity of the δ - barriers has been found to disappear at high sheet density of holes, > $6 \cdot 10^{16}$ m$^{-2}$, in Si-QWs as a result of the exchange interaction which leads to the formation of the spin polarons localized at the dipole boron centers.

The results obtained, specifically the linear decay of the magnetic susceptibility with increasing a magnetic field revealed by the *B-T* diagram in Fig. 10a at high temperature and in weak magnetic fields, have a bearing on the versions of the high temperature superconductivity that are based on the promising application of the sandwiches which consist of the alternating superconductor and insulator layers (Ginzburg 1964; Larkin and Ovchinnikov 1964; Fulde and Ferrell 1964; Little 1971). In the latter case, a series of heavily doped with boron and undoped silicon dots that forms the Josephson junction area in nanostructured δ - barriers is of advantage to achieve the high $T_c$ value, $T_c = (\hbar\omega_D/k_B)\exp(-N(0)V)$, because of the presence of the local high frequency phonon mode which compensates for relatively low density of states, *N(0)*.

Nevertheless, the mechanism of the bipolaronic transfer is still far from completely clear. This raises the question of whether the Josephson transitions dominate in the transfer of the pair of 2D holes in the plane of the nanostructured δ - barriers and in the proximity effect due to the tunneling through the Si-QW or the Andreev reflection plays a part in the bipolaronic transfer similar to the successive two-electron (hole) capture at the negative-U centers (Bagraev and Mashkov 1984; Bagraev and Mashkov 1988).

### 4. Superconducting proximity effect

Since the devices studied consist of a series of alternating semiconductor and superconductor nanostructures with dimensions comparable to both the Fermi wavelength and the superconductor coherence length, the periodic modulation of the critical current can be observed in consequence with quantum dimensional effects (Klapwijk 2004; van Dam et al. 2006; Jarillo-Herrero et al. 2006; Jie Xiang et al. 2006). Here the S-Si-QW-S structures performed in the Hall geometry are used to analyse the interplay between the phase-coherent tunneling in the normal state and the quantization of supercurrent in the superconducting state.

Firstly, the two-dimensional subbands of holes in the Si-QW identified by studying the far-infrared electroluminescence (EL) spectrum (Figs. 3a and b) appear to be revealed also by the I-V characteristic measured below the superconducting critical temperature of the δ-barriers which exhibits the modulation of the supercurrent flowing across the junction defined as the Josephson critical current (Fig. 15). The modulation of supercurrent seems to be caused by the tuning of on- and off-resonance with the subbands of 2D holes relatively to the Fermi energy in superconductor δ-barriers (Jarillo-Herrero et al. 2006; Jie Xiang et al. 2006) (see Figs. 16a and b). The two-dimensional subbands of 2D holes are revealed by varying the forward bias voltage (Figs. 15 and 16a), whereas the reverse bias voltage involves the levels that result from the Coulomb charging effects in the Si-QW filled with holes (Figs. 15 and 16b). The spectrum of supercurrent in the superconducting state appears to correlate with the conductance oscillations of the $2e^2/h$ value in the normal state of the S-Si-QW-S structure (Figs. 17a and b). This highest amplitude of the conductance oscillations is evidence of strong coupling in the superconductor δ-barriers (Fig. 17b). The data obtained demonstrate also that the amplitude of the quantum supercurrent is within frameworks of the well-known relationship $I_c R_n = \pi\Delta/e$ (Klapwijk 2004; Jie Xiang et al. 2006); where $R_n = 1/G_n$ is the normal resistance state, $2\Delta$ is superconducting gap, 0.044 eV. Besides, the strong coupling of on-resonance with the subbands of 2D holes which results from the $2e^2/h$ value of the conductance amplitude in the normal state is not related to the Kondo enhancement that is off-resonance (Cronenwett et al. 2002).

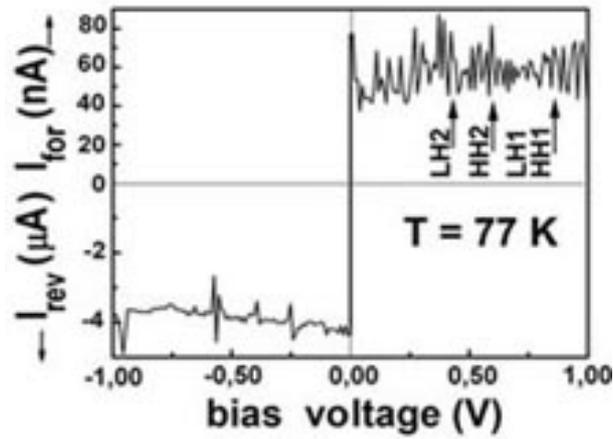

Fig. 15. *I-V* characteristic that demonstrates the modulation of the critical current with the forward and reverse bias applied to the p-type Si-QW confined by the δ-barriers on the n-type Si (100) surface.

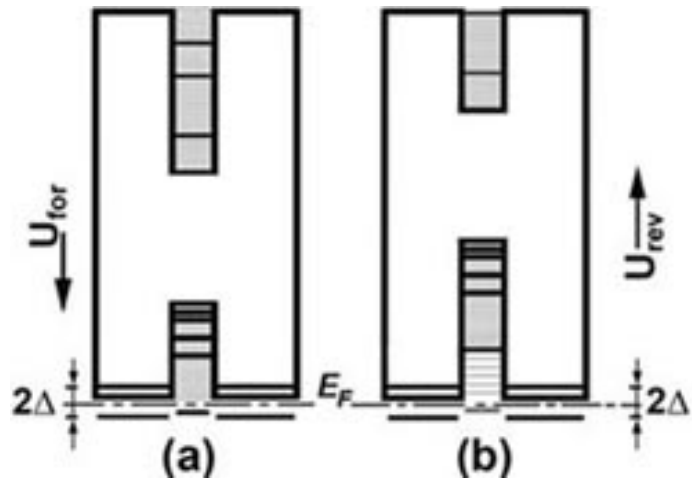

Fig. 16. The one-electron band scheme of the p-type Si-QW confined by the δ-barriers on the n-type Si (100) surface under forward (a) and reverse (b) bias, which depicts the superconducting gap, 2Δ, as well as the two-dimensional sub-bands of holes and the levels that result from the hole interference between the δ-barriers (a) and the Coulomb charging effects in the Si-QW filled with holes (b).

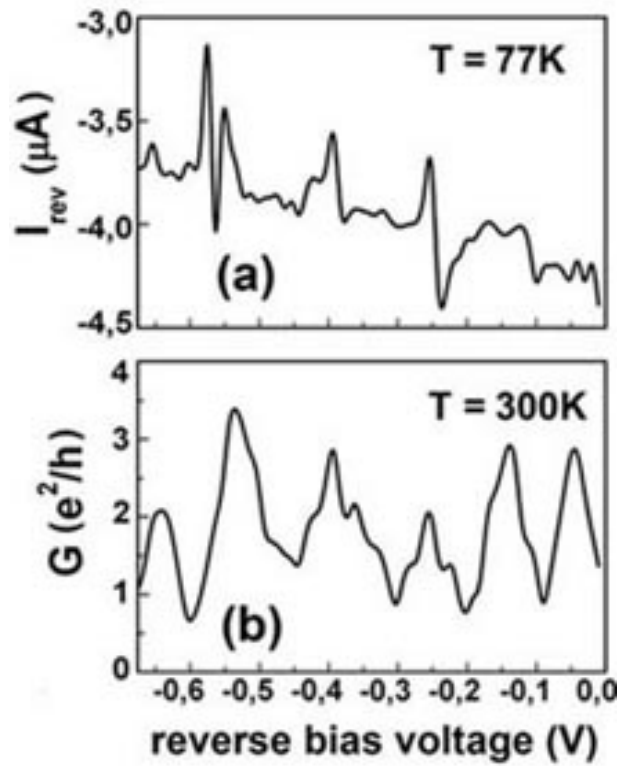

Fig. 17. Correlation between critical current (a) and normal state conductance (b) revealed by varying the reverse bias voltage applied to the sandwich structure, δ-barrier - p-type Si-QW - δ-barrier, on the n-type Si (100) surface.

Secondly, the spectrum of the supercurrent at low bias voltages appears to exhibit a series of peaks that are caused by multiple Andreev reflections (MARs) from the δ - barriers confining the Si-QW (Figs. 18a and b). The MAR process at the Si-QW - δ-barrier interface is due to the transformation of the 2D holes in a Cooper pair inside the superconducting δ-barrier which results in an electron being coherently reflected into the Si-QW, and vice versa, thereby providing the superconducting proximity effect (Figs. 19a and b) (Klapwijk 2004). The single hole crossing the Si-QW increases its energy by eV. Therefore, when the sum of these gains becomes to be equal to the superconducting energy gap, $2\Delta$, the resonant enhancement in the supercurrent is observed (Figs. 18a and b). The MAR peak positions occur at the voltages $V_n = 2\Delta/ne$, where $n$ is integer number, with the value $n=1$ related to the superconducting energy gap. It should be noted that the value of $2\Delta$, 0.033 eV, derived from the MAR oscillations does not agree with the magnetic susceptibility data because of heating of the device by bias voltage at finite temperatures. The mechanism of disappearance of some MAR peaks by varying the applied voltage is still in progress (Klapwijk 2004; Jarillo-Herrero et al. 2006; Jie Xiang et al. 2006). Nevertheless, the linear dependence of the MAR peak position on the value of $1/n$ was observed.

The MAR processes are of interest to be measured in the regime of coherent tunneling (Eisenstein et al. 1991) in the studies of the device performed in frameworks of the Hall geometry, because the phase coherence is provided by the Andreev reflection of the single holes (electrons) at the same angle relatively to the Si-QW plane. In the device studied this angle is determined by the crystallographic orientation of the trigonal dipole centers of boron inside the δ-barriers (Figs. 6a and b). These MAR processes were observed as the oscillations of the longitudinal conductance by varying the value of the top gate voltage, with the linear dependence of the MAR peak position on the value of $1/n$ (Bagraev et al. 2008c). The value of the superconducting energy gap, 0.044 eV, derived from these dependences was in a good agreement with the magnetic susceptibility data that is evidence of the absence of heating effects at the values of the drain-source voltage used in the regime of coherent tunneling. The amplitude of the MAR peaks observed, $e^2/h$, appeared to be independent of the value of the drain-source voltage that is also attributable to the coherent tunneling. Since the MAR processes are spin-dependent (Klapwijk 2004), the effect of the Rashba SOI created at the same geometry by varying the value of the top gate voltage appears to be responsible for the mechanism of the coherent tunneling in Si-QW. In addition to the $e^2/h$ amplitude of the MAR peaks, this concept seems to result from the stability of the Fermi wave vector that was controlled in the corresponding range of the top gate voltage by the Hall measurements. Within frameworks of this mechanism of the coherent tunneling, the spin projection of 2D holes that take part in the MAR processes is conserved in the Si-QW plane (Klapwijk 2004) and its precession in the Rashba effective field is able to give rise to the reproduction of the MAR peaks in the oscillations of the longitudinal conductance. Thus, the interplay of the MAR processes and the Rashba SOI appears to reveal the spin transistor effect (Bagraev et al. 2005; Bagraev et al. 2006a; Bagraev et al. 2008a; Bagraev et al. 2008c) without the injection of the spin-polarized carriers from the iron contacts as proposed in the classical version of this device.

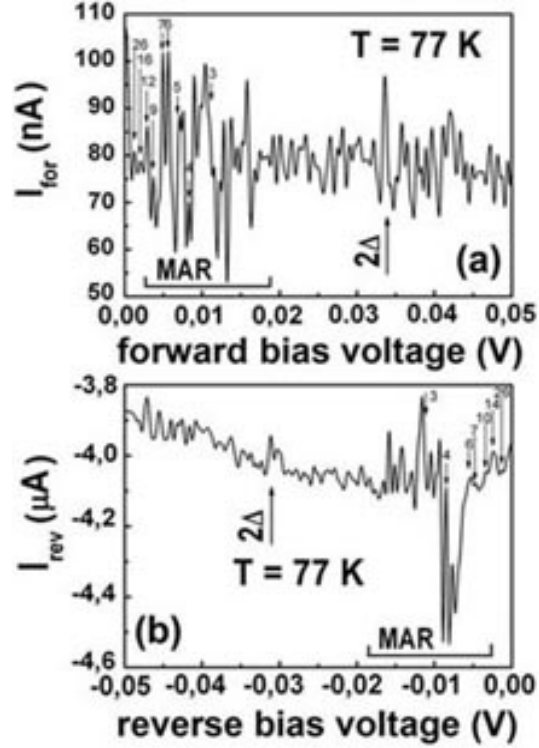

Fig. 18. Multiple Andreev reflections (MAR) with the forward (a) and reverse (b) bias applied to the sandwich structure, δ-barrier - p-type Si-QW - δ-barrier on the n-type Si (100) surface. The MAR peak positions are marked at $V_n = 2\Delta/ne$ with values $n$ indicated. The superconducting gap peak, $2\Delta$, is also present. The difference in the values of critical current under forward and reverse bias voltage is due to non-symmetry of the sandwich structure.

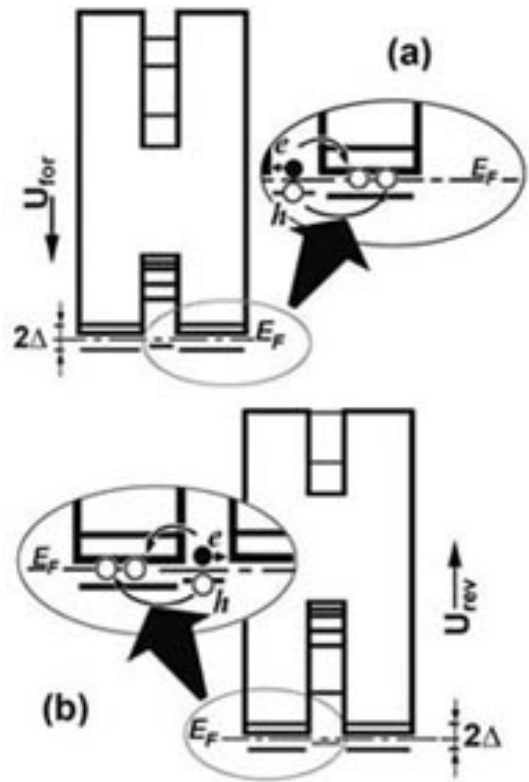

Fig. 19. The one-electron band scheme of the sandwich structure, δ-barrier - p-type Si-QW - δ-barrier, on the n-type Si (100) surface that reveals the multiple Andreev reflection (MAR) caused by pair hole tunneling into δ-barrier under forward (a) and reverse (b) bias.

Finally, the studies of the proximity effect in the 'sandwich' S-Si-QW-S structures have shown that the MAR processes are of great concern in the transfer of the small hole bipolarons both between and along nanostructured δ-barriers confining the Si-QW. Within MAR processes, the pairs of 2D holes introduced into the δ-barriers from the Si-QW seem to serve as the basis for the bipolaronic transfer that represents the successive coherent tunneling of small hole bipolarons through the dipole boron centers up the point, of which an electron is coherently reflected into the Si-QW. The most likely tunneling through the negative-U centers appear to be due to the successive capture of two holes accompanied by their generation or single-electron emission in consequence with the Auger processes:

$B^+ + B^- + 2h \Rightarrow B^+ + B^0 + h \Rightarrow B^0 + B^+ + h \Rightarrow B^- + B^+ + 2h$ or $B^+ + B^- \Rightarrow B^+ + B^0 + e \Rightarrow B^0 + B^0 + h + e \Rightarrow B^- + B^0 + 2h + e \Rightarrow B^- + B^+ + h + e$. Relative contribution of these processes determines the coherence length. Besides, the single-hole tunnelling through the negative-U centers that is able to increase the critical temperature should be also taken into account (Šimánek 1979; Ting et al. 1980). Thus, the charge and spin density waves seem to be formed along the δ-barrier – Si-QW interface with the coherence length defined by the length of the bipolaronic transfer that is dependent on the MAR characteristics.

**Summary**

Superconductivity of the sandwich' S-Si-QW-S structures that represent the p-type high mobility silicon quantum wells confined by the nanostructured δ - barriers heavily doped with boron on the n-type Si (100) surface has been demonstrated in the measurements of the temperature and magnetic field dependencies of the resistance, thermo-emf, specific heat and magnetic susceptibility.

The studies of the cyclotron resonance angular dependences, the scanning tunneling microscopy images and the electron spin resonance have shown that the nanostructured δ - barriers consist of a series of alternating undoped and doped quantum dots, with the doped dots containing the single trigonal dipole centers, $B^+ - B^-$, which are caused by the negative-U reconstruction of the shallow boron acceptors, $2B^0 \rightarrow B^+ + B^-$.

The temperature and magnetic field dependencies of the resistance, thermo-emf, specific heat and magnetic susceptibility are evidence of the high temperature superconductivity, $T_c = 145$ K, that seems to result from the transfer of the small hole bipolarons through these negative-U dipole centers of boron at the Si-QW – δ - barrier interfaces.

The oscillations of the upper critical field and critical temperature vs magnetic field and temperature that result from the quantization of the critical current have been found using the specific heat and magnetic susceptibility techniques.

The value of the superconductor energy gap, 0.044 eV, derived from the measurements of the critical temperature using the different techniques appeared to be practically identical to the data of the current-voltage characteristics and the local tunneling spectroscopy.

The extremely low value of the hole effective mass in the 'sandwich' S-Si-QW-S structures that has been derived from the measurements of the SdH oscillations seems to be the principal argument for the bipolaronic mechanism of high temperature superconductor properties that is based on the coherent tunneling of bipolarons. The high frequency local phonon mode that is revealed with the superconductor energy gap in the infrared transmission spectra seems also to be responsible for the formation and the transfer of small hole bipolarons.

The proximity effect in the S-Si-QW-S structure has been identified by the findings of the MAR processes and the quantization of the supercurrent. The value of the superconductor energy gap, 0.044 eV, appeared to be in a good agreement with the data derived from the oscillations of the conductance in normal state and of the zero-resistance supercurrent in superconductor state as a function of the bias voltage. These oscillations have been found to be correlated by on- and off-resonance tuning the two-dimensional subbands of holes with the Fermi energy in the superconductor δ - barriers.

Finally, the studies of the proximity effect in the 'sandwich' S-Si-QW-S structures have shown that the multiple Andreev reflection (MAR) processes are of great concern in the coherent transfer of the small hole bipolarons both between and along nanostructured δ-barriers confining the Si-QW.

**Acknowledgements**

The work was supported by the SNSF programme (grant IB7320-110970/1), RAS-QN (grant 4-2. 9A-19), RAS-QM (grant P-03. 4.1).

32Slaoui, A., Fogarassy, E., Muller, J.C., Siffert, P.: Study of some optical and electrical properties of heavily doped silicon layers. J. de Physique Colloq. C5 **44**, 65-71 (1983)

de Souza Silva, C.C., van de Vondel, J., Morelle, M., Moshchalkov, V.V.: Controlled multiple reversals of ratchet effect. Nature **440**, 651-654 (2006)

Street, R.A., Mott, N.F.: States in the gap in glassy semiconductors. Phys. Rev. Lett. **35**, 1293 - 1296 (1975)

Suderow, H., Bascones, E., Izquierdo, A., Guinea, F., Vieira, S.: Proximity effect and strong-coupling superconductivity in nanostructures built with an STM. Phys. Rev. B **65**, 100519-4 (2002)

Šimánek, E.: Superconductivity at disordered interfaces. Solid State Comm. **32**, 731–734 (1979)

Ting, C.S., Talwar, D.N., Ngai, K.L.: Possible mechanism of superconductivity in metal-semiconductor eutectic alloys. Phys. Rev. Lett. **45**, 1213–1216 (1980)

Trovarelli, O., Weiden, M., Müller-Reisener, R., Gómez-Berisso, M., Gegenwart, P., Deppe, M., Geibel, C., Sereni, J.G., Steglich, F.: Evolution of magnetism and superconductivity in $CeCu_2(Si_{1-x}Ge_x)_2$. Phys. Rev. B **56**, 678–685 (1997)

Vodolazov, D.Y., Golubović, D.S., Peeters, F.M., Moshchalkov, V.V.: Enhancement and decrease of critical current due to suppression of superconductivity by a magnetic field. Phys. Rev. B **76**, 134505-7 (2007)

Watkins, G.D.: Negative-U properties for defects in solids, Festkoerperprobleme, **24**, 163-186 (1984)

Weisbuch, C., Vinter, B.: Quantum semiconductor structures. Academic Press, Boston (1991)

Werthamer, N.R., Helfand, E., Hohenberg, P.C.: Temperature and purity dependence of the superconducting critical field $H_{c2}$. III. Electron spin and spin-orbit effects. Phys. Rev. **147**, 295–302 (1966)